\documentclass[%
twocolumn,
 amsmath,amssymb,
 aps,
pra,
]{revtex4-2}
\usepackage{orcidlink}
\usepackage{graphicx}
\usepackage{physics}
\usepackage{dcolumn}
\usepackage{bm}
 
\usepackage{blkarray}
\usepackage{amsmath,calc}

\begin{document}
\preprint{APS/123-QED}

\title{Perturbing finite temperature multicomponent DFT 1D Kohn-Sham systems: \\Peierls Gap \& Kohn Anomaly}

\author{Adrian D Scheppe\orcidlink{0000-0002-5566-2744}}\email{adrian.scheppe@afit.edu}
\author{Michael V. Pak}\email{michael.pak@afit.edu}
\affiliation{Department of Physics, Air Force Institute of Technology\\
2950 Hobson Way, Wright-Patterson AFB, Ohio 45433}
\date{10 November 2023}

\begin{abstract}
One of the greatest challenges when designing new technologies that make use of non-trivial quantum materials is the difficulty associated with predicting material-specific properties, such as critical temperature, gap parameter, etc. There is naturally a great amount of interest in these types of condensed matter systems because of their application to quantum sensing, quantum electronics, and quantum computation; however, they are exceedingly difficult to address from first principles because of the famous many-body problem. For this reason, a full electron-nuclear quantum calculation will likely remain completely out of reach for the foreseeable future. A practical alternative is provided by finite temperature, multi component density functional theory (MCDFT), which is a formally exact method of computing the equilibrium state energy of a many-body quantum system. In this work, we use this construction alongside a perturbative scheme to demonstrate that the phenomena Peierls effect and Kohn Anomaly are both natural features of the KS equations without additional structure needed. We find the temperature dependent ionic density for a simple 1D lattice which is then used to derive the ionic densities temperature dependent affect on the electronic band structure. This is accomplished by Fourier transforming the ionic density term found within this KS electronic equation. Using the Peierls effect phonon distortion gap openings in relation to the Fermi level, we then perturb the KS ionic equation with a conduction electron density, deriving the Kohn Anomaly. This provides a workable predictive strategy for interesting electro-phonon related material properties which could be extended to 2D and 3D real materials while retaining the otherwise complicated temperature dependence.
\end{abstract}

\maketitle
\section{Introduction}
Somewhere near the top of the lengthy catalogue of critical $20^{\text{th}}$ century discoveries sits the observation of superconductivity, a property that emerges due to the effective pairing interaction of electrons and the subsequent formation of Cooper pairs \cite{hott2013review, sato2017topological}. It is widely known that the condensate that arises from this reformulation of the ground state is ultimately responsible for the mysterious behavior exhibited by such systems, and it is due to these qualities, that such materials find themselves at the center of several astonishing quantum technological fetes such as regular and fault tolerant qubits \cite{krantz2019quantum,scheppe2022complete,schrade2018majorana}, phase engineered Josephson Junctions \cite{strambini2016omega}, topological heterostructure electronics \cite{gilbert2021topological}, and so on. So, it seems currently that the superconductor (SC) will form the backbone of technology leading into the future, and, for this reason, we are sufficiently motivated to streamline our search and synthesis of such materials.

In an experimental setting, it is possible to determine when a material is trivial or not, and the list of discovered materials grows every year \cite{hott2013review, sharma2022comprehensive}. Crucially though, the act of making a judgement of a material's nature \textit{prior} to experiment is not so straightforward, and, so far, seems impossible. Current powerful predictive machine learned algorithms have the ability to accurately guess critical temperature for many materials \cite{dan2020computational,claussen2020detection,xie2019functional,stanev2018machine,zhang2021predicting,campi2021prediction,matasov2020prediction,le2020critical}, and, amazingly, they have even predicted that structures would exhibit SC nature before the material had been synthesized \cite{gou2013discovery}. However, these methods carry the burden of requiring extremely large data sets which are ultimately sourced by experimentation and have difficulty predicting material properties as well \cite{malyi2020realization}.

The reason for guessing is that there is a significant knowledge gap in regard to these materials. The usual quantum mechanical treatment of any system begins with the infamous many-body Hamiltonian which is known to be exceedingly difficult, requiring a series of simplification strategies to obtain a solution. Naturally, simplifications are a necessary evil which whittle away the accuracy of the model in an attempt to derive a workable solution. However, in this case, typical strategies obfuscate the true nature of the system, meaning the conventional approaches are completely blind to superconductivity. To make predictions then, one reintroduces an attractive electron field, $\Delta(\textbf{r},\textbf{r}')$, as a new degree of freedom to the simplified Hamiltonian. The explicit inclusion of such a term works very well if we have prior knowledge that the system will exhibit SC nature; however, the procedure is not so straight forward when dealing with an unknown material. This is not to say that this is a poor method with bad predictions, but rather we say the opposite that this approach is unpredictable in how it fails. In this way, we do not have predictive knowledge of a material's non-trivial status and are forced to slowly probe each material one-by-one. This hampers the development of any quantum technology in general. 

Developments in Finite Temperature Multi Component Density Functional Theory (MCDFT) offer an alternative, formally exact technique which simplifies the many-body problem, replacing the direct interactions of each subsystem with statistically averaged electronic and ionic densities \cite{van2003density,pribram2014thermal,kreibich2001multicomponent}. Leading approaches which make use of this framework explicitly include a pairing field density within the model Hamiltonian alongside the electronic and ionic densities \cite{oliveira1988density,luders2005ab}. The critical temperature and gap parameter obtained via this method are in good agreement for the elemental SC, and, naturally, not as good for more complicated materials \cite{continenza2005ab}. These methods are a crucial step for the field of DFT and the search for non-trivial materials; however, we wish to supplement this scheme with an investigation into the nature of the Kohn-Sham (KS) equations which may already possess the fidelity required to derive SC nature without the explicit inclusion of a pairing field or pairing density. This method provides a suitable starting point to reassess the many-body problem and provides the possible key to deriving electro-phonon properties directly from the many-body Hamiltonian.

To this end, we derive qualities associated with the electro-phonon subsystems from the KS equations using simple qualitative arguments made from the assumption of phonon mediation. We apply the two KS equations to a 1D lattice which retains a temperature description within the statistically averaged density of the ions related to Bose-Einstein phonon statistics. From this we derive a general Peierls effect for any phonon wave number as a natural temperature dependent feature of the combined system. Additionally, we derive the Kohn anomaly via perturbative means as a conduction electron density reaction to the phonon statistical gap opening. These two phenomena are the markers of interesting behavior associated with SC materials which \textit{naturally} fall out of the KS equations as a direct consequence of a relationship between the two subsystems, demonstrating that this method of simplification possesses sufficient fidelity to predict interesting electro-phonon behaviour. The full extent of the relationship between the Kohn anomaly and SC is still unknown, but it is clear that there is a relationship in general \cite{chaudhury2013kohn, chaudhury2010kohn, aynajian2008energy}. From this, we speculate that the KS equations within this method could be able to show that a material will naturally superconduct.

This document is written in such a way to be as self contained as possible. We first outline qualities of electronic and phononic subsystems in Sect. \ref{sec:SClogic} which must be true to facilitate the logical flow of the work. Sect. \ref{sec:MCDFT} covers the derivation of the MCDFT scheme and how one arrives at the KS equations. Then, in Sect. \ref{sec:1Dchain}, we apply the scheme to a 1D chain of ions to derive the phonon effect on the electronic band structure and vice versa.

\section{Qualitative SC Phase Transition}
\label{sec:SClogic}
Prior to our theoretical considerations, it is necessary to state the qualitative interaction between the two subsystems. Assuming the ideas behind phonon mediated SC phase transition are correct, we can at the very least speak qualitatively about what occurs on a microscopic level to facilitate SC. Here is a list of guided facts which are meant to facilitate the logical flow of the rest of the document:
\begin{enumerate}
    \item SC is a bulk property that emerges due to the electron-phonon interaction, meaning our simplified model must retain some key aspect of the fundamental interactions between the bosonic phonons and the fermionic electrons.
    \item Individual electrons are not large enough to sway the trends of the lattice vibrations, and they do not \textit{feel} temperature, i.e. the statistical mechanics of the electrons are essentially ground state for our range of temperatures.
    \item On the other hand, phonons \textit{do} feel temperature, so it is reasonable then to assume that the lattice instigates the initial transition to SC phase as $T\rightarrow T_C$. The electronic densities reaction to this transition come from not the lowering temperature exactly, but from a trend in the phonon occupation which constrains the possible electronic densities.
    \item The electronic densities represent large collections of electrons which \textit{do} have the ability to sway the lattice. They somehow fall into a configuration which encourages the phonon trend that initiated the process.
    \item The system falls into a resonance pattern which is energetically favorable, yielding the markers of SC, i.e cooper pairs, Meissner effect, persistent current, etc. 
\end{enumerate}
Using this line of thinking, we expect there to be a deviation from what statistical mechanics predicts for a Debye solid at the critical temperature.
\section{Finite Temperature MCDFT}
\label{sec:MCDFT}
For absolute clarity, this section is provided for non Density Functional theorists; however, to understand the following sections, it is not at all necessary for the reader to understand how to arrive at the KS equations. This section takes its material from Refs. \cite{van2003density, pribram2014thermal,luders2005ab,kreibich2001multicomponent,mermin1965thermal,hohenberg1964inhomogeneous,butriy2007multicomponent}. 
Specifically, we follow the finite temperature MCDFT methods outlined within Ref. \cite{luders2005ab} which assume the existence of a Hohenberg-Kohn type relationship between external potentials and densities even in the finite temperature regime. This assumption is plausible but not proven, so long as one neglects the Mass Polarization-Coriolis terms which appear as a result of transforming to a body fixed coordinate system \cite{kreibich2001multicomponent}. However, it is important to note here that it is exactly this aspect of MCDFT that makes it difficult to prove a finite temperature extension to Hohenberg-Kohn theorem. This is due to the fact that such a transformation is dependent on the number of particles, and therefore different for each state in the ensemble. Nonetheless, we wish to apply the established formalism with the understanding that full justification awaits.

\subsection{Grand Canonical Energy Functional}
Starting with the abstract many-body Hamiltonian,
\begin{equation*}
    \hat{H}=\hat{T}^e+\hat{T}^i+\hat{W}^{ee}+\hat{W}^{ii}+\hat{W}^{ei},
\end{equation*}
where $\hat{T}$ and $\hat{W}$ denote kinetic energies and interaction potentials for electronic, $e$, and ionic, $i$, subsystems. The statistical mechanics enter the picture as ensemble constraints placed upon the Hamiltonian. The Grand Canonical ensemble allows for changes in the particle number and energy of each ensemble member where the average particle number, average energy, and number of ensembles lead to chemical potential, temperature, and normalization (partition function) respectively. The former two enter via Lagrange multipliers $\mu$ and $\tau$,
\begin{equation*}
    \hat{\Omega}=\hat{H}-\mu \hat{N}+\tau\hat{S},
\end{equation*}
where $\hat{N}$ is the number operator for electrons, and $\hat{S}$ is von Neumann entropy,
\begin{align*}
    \hat{N}&=\sum_{\sigma}\Psi_{\sigma}^\dagger(\textbf{r})\Psi_{\sigma}(\textbf{r}),\\
    \hat{S}&=-\text{Tr}\{\rho \text{ln}(\rho)\},
\end{align*}
where $\Psi_{\sigma}(\textbf{r})$ is electronic field operator. The projection operator $\rho$ is already known to be the Grand Canonical ensemble,
\begin{equation*}
\rho_0=\frac{1}{\text{Tr}\{e^{\beta(\hat{H}-\mu \hat{N})}\}}e^{\beta(\hat{H}-\mu \hat{N})},
\end{equation*}
because of statistical mechanics minimization. This minimization procedure precludes the functional minimization procedure on the DFT side of things, so this operator is left fixed throughout the MCDFT theory.

With that, we assume a MCDFT Hohenberg-Kohn type statement exists\cite{hohenberg1964inhomogeneous,luders2005ab}, i.e,
\begin{align*}
    \text{1-to-1 Map: }&\{n(\textbf{r}),\Gamma(\underline{\textbf{R}})\}\leftrightarrow\{v^e_{ext}(\textbf{r}),v^i_{ext}(\underline{\textbf{R}})\},\\
    \text{Observables: }&\langle\mathcal{O}\rangle=\mathcal{O}[n(\textbf{r}),\Gamma(\underline{\textbf{R}})],\\
    \text{Variation: }&\mathcal{O}_{eq}\le\mathcal{O}[n(\textbf{r}),\Gamma(\underline{\textbf{R}})].
\end{align*}
For now, the definitions of densities and external potentials are left ambiguous intentionally.

A strict 1-to-1 link between the set of defined densities and external potentials mean that the two sets are essentially the same thing and contain the same information. This fact is important because it shows that we need to be careful when defining our external potentials. In regular electronic DFT, the external potential is justifiably the lattice potential and true external potentials (i.e. voltage bias, external fields, etc.). Treating the ionic interaction as an external potential is crucial because otherwise, the 1-to-1 map guarantee would force the density to be constant, showing no internal signature of the system in question. In the case of MCDFT, the situation is more complicated because we have two densities, electronic and ionic, but the motivation is the same. 

For this reason, we don't define densities freely. We first need to pick which terms in $\hat{\Omega}$ that are to be promoted to the external potential. These are conjugate to the densities, and therefore facilitate the establishment of density definitions. Following Ref. \cite{luders2005ab}, we choose the ionic interaction and chemical potential terms to be external,
\begin{align*}
    v^e_{ext}(\textbf{r})&=-\mu,\\
    v^i_{ext}(\underline{\textbf{R}})&=\frac{1}{2}\sum_{\alpha\beta}\frac{Z^2}{\abs{R_\alpha-R_\beta}}.
\end{align*}
These potentials are then conjugate to the set of densities,
\begin{align*}
    n(\textbf{r})&=\sum_\sigma\langle \Psi^\dagger_\sigma(\textbf{r})\Psi_\sigma(\textbf{r})\rangle,\\
    \Gamma(\underline{\textbf{R}})&=\langle \Phi^\dagger(\textbf{R}_1)...\Phi^\dagger(\textbf{R}_N)\Phi(\textbf{R}_N)...\Phi(\textbf{R}_1) \rangle,
\end{align*}
where $\langle...\rangle$ is the expectation value with respect to $\rho_0$.
Using these definitions, we can now write out the general form of the Grand Canonical energy functional,
\begin{align*}
\Omega[n(\textbf{r}),\Gamma(\underline{\textbf{R}})]=F[n(\textbf{r}),\Gamma(\underline{\textbf{R}})]+\int d\textbf{r}n(\textbf{r})&v^e_{ext}(\textbf{r})\\
+\int d\underline{\textbf{R}}&\Gamma(\underline{\textbf{R}})v^i_{ext}(\underline{\textbf{R}}),
\end{align*}
where the universal functional is defined like so,
\begin{align*}
F[n(\textbf{r}),\Gamma(\underline{\textbf{R}})]=T^e[n(\textbf{r})]+T^i[\Gamma(\underline{\textbf{R}})]+&W^{ee}[n(\textbf{r})]\\+W^{ii}[\Gamma(\underline{\textbf{R}})]+W^{ei}[n(\textbf{r}),\Gamma(\underline{\textbf{R}})]&-\tau S[n(\textbf{r}),\Gamma(\underline{\textbf{R}})].
\end{align*}
We can see the link between density and potential on full display here by taking the functional derivative with respect to each density,
\begin{align}
    \frac{\delta\Omega}{\delta n(\textbf{r})}=\frac{\delta F}{\delta n(\textbf{r})}+v^e_{ext}(\textbf{r})=0\hspace{2pt}&\rightarrow\hspace{2pt}\frac{\delta F}{\delta n(\textbf{r})}=-v^e_{ext}(\textbf{r}),\label{eq:legTrans1}\\
    \vspace{4pt}\nonumber\\
    \frac{\delta\Omega}{\delta \Gamma(\underline{\textbf{R}})}=\frac{\delta F}{\delta \Gamma(\underline{\textbf{R}})}+v^i_{ext}(\textbf{\underline{R}})=0\hspace{2pt}&\rightarrow\hspace{2pt}\frac{\delta F}{\delta \Gamma(\underline{\textbf{R}})}=-v^i_{ext}(\textbf{\underline{R}}),\label{eq:legTrans2}
\end{align}
such that,
\begin{align*}
\Omega[n(\textbf{r}),\Gamma(\underline{\textbf{R}})]=F[n(\textbf{r}),\Gamma(\underline{\textbf{R}})]-\int &d\textbf{r}n(\textbf{r})\frac{\delta F}{\delta n(\textbf{r})}\\-&\int d\underline{\textbf{R}}\Gamma(\underline{\textbf{R}})\frac{\delta F}{\delta \Gamma(\underline{\textbf{R}})}.
\end{align*}
This is simply the functional version of a Legendre transform between two conjugate functions, densities and potentials \cite{van2003density}. 
\subsection{Thermal MCDFT Kohn-Sham Scheme}
The KS scheme simplification procedure replaces the direct interaction terms of each subsystem with statistically averaged electronic and ionic density interaction. First, assume the non-interacting ($W^{ee}=W^{ei}=0$) KS system shares the same density as the fully interacting one. This system has its own energy and universal functionals,
\begin{align*}
\Omega_s[n(\textbf{r}),\Gamma(\underline{\textbf{R}})]=F_s[n(\textbf{r}),\Gamma(\underline{\textbf{R}})]+&\int d\textbf{r}n(\textbf{r})\frac{\delta F_s}{\delta n(\textbf{r})}\\&+\int d\underline{\textbf{R}}\Gamma(\underline{\textbf{R}})\frac{\delta F_s}{\delta \Gamma(\underline{\textbf{R}})},
\end{align*}
\begin{align*}
F_s[n(\textbf{r}),\Gamma(\underline{\textbf{R}})]=T_s^e[n(\textbf{r})]+T_s^i[\Gamma(\underline{\textbf{R}})]&+W^{ii}_s[\Gamma(\underline{\textbf{R}})]\\&-\tau S_s[n(\textbf{r}),\Gamma(\underline{\textbf{R}})].
\end{align*}
Now, making the standard definition of exchange correlation, i.e the difference between fully interacting and KS systems with the Hartree terms replacing the terms set to zero, one can restate the universal functional as,
\begin{align*}
F[n(\textbf{r}),\Gamma(\underline{\textbf{R}})]=F_s[n(\textbf{r}),&\Gamma(\underline{\textbf{R}})]+F_{xc}[n(\textbf{r}),\Gamma(\underline{\textbf{R}})]\\&+E^{ee}[n(\textbf{r})]+E^{ei}[n(\textbf{r}),\Gamma(\underline{\textbf{R}})],
\end{align*}
where,
\begin{align*}
    E^{ee}[n(\textbf{r})]&=\frac{1}{2}\int d\textbf{r}d\textbf{r}'\frac{n(\textbf{r})n(\textbf{r}')}{\abs{\textbf{r}-\textbf{r}'}}\\
    E^{ei}[n(\textbf{r}),\Gamma(\underline{\textbf{R}})]&=\sum_{\alpha}\int d\textbf{r}d\underline{\textbf{R}}\frac{n(\textbf{r})\Gamma(\underline{\textbf{R}})}{\abs{\textbf{r}-\textbf{R}_\alpha}},
\end{align*}
are the Hartree terms corresponding to each interaction set to zero in the KS system.

We have the freedom to vary the Grand potential with respect to $n$ and $\Gamma$ independently, so, using Eq. \ref{eq:legTrans1} and Eq. \ref{eq:legTrans2},
\begin{align*}
    \frac{\delta F}{\delta n(\textbf{r})}=\frac{\delta F_s}{\delta n(\textbf{r})}+\frac{\delta F_{xc}}{\delta n(\textbf{r})}+\frac{\delta E^{ee}}{\delta n(\textbf{r})}+\frac{\delta E^{ei}}{\delta n(\textbf{r})}=-v^e_{ext}(\textbf{r}),
\end{align*}
\begin{align*}
    \frac{\delta F}{\delta \Gamma(\underline{\textbf{R}})}=\frac{\delta F_s}{\delta \Gamma(\underline{\textbf{R}})}+\frac{\delta F_{xc}}{\delta \Gamma(\underline{\textbf{R}})}+&\frac{\delta E^{ee}}{\delta \Gamma(\underline{\textbf{R}})}\\&+\frac{\delta E^{ei}}{\delta \Gamma(\underline{\textbf{R}})}=-v^i_{ext}(\textbf{\underline{R}}).
\end{align*}
The KS potential can then be written out for each subsystem. For the electronic system,
\begin{align*}
    v^e_{s,ext}(\textbf{r})=v^e_{xc}[n,\Gamma](\textbf{r})+\int d&\textbf{r}'\frac{n(\textbf{r}')}{\abs{\textbf{r}-\textbf{r}'}}\\
    &+\sum_{\alpha}\int d\underline{\textbf{R}}\frac{\Gamma(\underline{\textbf{R}})}{\abs{\textbf{r}-\textbf{R}_\alpha}}-\mu,
\end{align*}
where we make use of the conjugate relationship for the KS system analogous to the one in Eqs. \ref{eq:legTrans1} and \ref{eq:legTrans2}. Additionally, the functional derivatives have been evaluated and the external potential has been used. This establishes the electronic KS potential where the entire expression is an external potential to non-interacting Bloch electrons. We do the same for the ionic system,
\begin{align*}
    v^i_{s,ext}(\underline{\textbf{R}})=v^i_{xc}[n,\Gamma](\underline{\textbf{R}})+\sum_{\alpha}&\int d\textbf{r}\frac{n(\textbf{r})}{\abs{\textbf{r}-\textbf{R}_\alpha}}\\
    &+\frac{1}{2}\sum_{\alpha\beta}\frac{Z^2}{\abs{R_\alpha-R_\beta}},
\end{align*}
yielding the ionic KS equation. We can now write out the KS equations,
\begin{align}
    &\{\frac{1}{2m_e}\nabla^2+v^e_{s,ext}(\textbf{r})-\mu\}\phi_{i,\sigma}(\textbf{r})=\varepsilon_i\phi_{i,\sigma}(\textbf{r}),\label{eq:ksEqnE}\\
    &\{\frac{1}{2M}\sum_\alpha\nabla_\alpha^2+v^i_{s,ext}(\underline{\textbf{R}})\}\Phi(\underline{\textbf{R}})=E\Phi(\underline{\textbf{R}}).\label{eq:ksEqnN}
\end{align}
Thereby completing the process of simplifying the original Hamiltonian into a manageable alternative set of equations.
\subsection{Densities}
One can write a simpler, general form for each density using the KS assumption. For non-interacting, Bloch electrons,
\begin{equation*}
    n(\textbf{r})=\sum_\sigma\text{Tr}\{\rho_0\phi^\dagger_\sigma(\textbf{r})\phi_\sigma(\textbf{r})\}=\sum_i g^\tau_i\abs{\phi_i(\textbf{r})}^2,
\end{equation*}
where $g^\tau_i$ is the Fermi-Dirac distribution and $\phi_i(\textbf{r})$ are non-interacting Bloch states. It is at this point we can capitalize on one of the statements within Sect. \ref{sec:SClogic}. Since electrons are largely in the ground state for all temperatures we are concerned with, we can treat the KS Bloch electron density as ground state, approximating,
\begin{align*}
    n(\textbf{r})\approx\sum^{Occ}_{i}\abs{\phi_i(\textbf{r})},
\end{align*}
which sums over all occupied states up to the Fermi level. This sum is particularly easy to address in our case where we are interested in picking out a Fermi level electronic density. For a more accurate problem, one would integrate over a density of states keeping the bounds of integration close to the Fermi level.

The ionic density is slightly more tricky because of the ionic collective state, but the exact procedure for handling this is well covered in Ref. \cite{gross1986many, ashcroft2022solid}. Starting with the definition,
\begin{equation*}
    \Gamma(\underline{\textbf{R}})=\langle \Phi^\dagger(\textbf{R}_1)...\Phi^\dagger(\textbf{R}_N)\Phi(\textbf{R}_N)...\Phi(\textbf{R}_1) \rangle,
\end{equation*}
we make the necessary coordinate transformation on the classical phase space to normal mode coordinates,
\begin{equation*}
    \{\textbf{R}_1,...,\textbf{R}_N\}\rightarrow\{\textbf{Q}_1,...,\textbf{Q}_N\},
\end{equation*}
which induces a transformation on the states with respect to these coordinates,
\begin{equation*}
    \Phi^\dagger(\textbf{R}_1)...\Phi^\dagger(\textbf{R}_N)\rightarrow h_n(Q_{\textbf{k},\lambda}).
\end{equation*}
Under the harmonic approximation, the ionic state with respect to the normal mode coordinates is a simple harmonic oscillator, $h_n(Q_{\textbf{k},\lambda})$, where $\textbf{k}$ is the phonon wave number, $\lambda$ labels the normal mode, and $n$ labels the energy level. The creation operator for this state is now a single Bosonic creation operator, $\hat{b}^\dagger_{\textbf{k}}$. This subsequently transforms the original density like so,
\begin{equation}
\label{eq:gammaGeneral}
    \Gamma(\underline{\textbf{Q}})=\sum_{n,\textbf{k},\lambda}f^\tau_{n,\textbf{k},\lambda}\abs{h_n(Q_{\textbf{k},\lambda})}^2,
\end{equation}
where $f^\tau_{n,\textbf{k},\lambda}$ is the Bose-Einstein distribution. This distribution assigns a statistical weight to each $\textbf{k}$-vector (wave direction) and each $\lambda$-mode (style of vibration). Each harmonic oscillator that sits at every point in the BZ is summed together where modes with higher occupation numbers have more say in what the lattice is doing. We note here that transverse (shear) waves do not cause any appreciable change to the positive density that electrons scatter off of; however, the longitudinal (compression) waves do in fact change the positive density and the band structure problem \cite{gross1971many}. 

\section{1D Model:}
\label{sec:1Dchain}
The goal in this section is to use the 1D chain as a demonstration of what happens when the lattice becomes selective in its oscillations. This result for the 1D chain of oscillators is already known as Peierls transition where distortions in the lattice alter the electronic band structure problem \cite{gutfreund1974peierls, horovitz1977peierls}. There are, however, a few items missing from the typical Peierls gap treatment. How the lattice becomes selective in its distortion is not captured in the literature regarding this phenomenon. It is known that this phenomenon is linked to temperature; however, the works outlining the Peierls phenomenon and the gap opening simply state that the lattice becomes spontaneously distorted at some temperature, beginning calculations with a ``frozen" phonon which changes the lattice periodicity.

\begin{figure}[t!]
    \centering
    \includegraphics[width = 
    \linewidth]{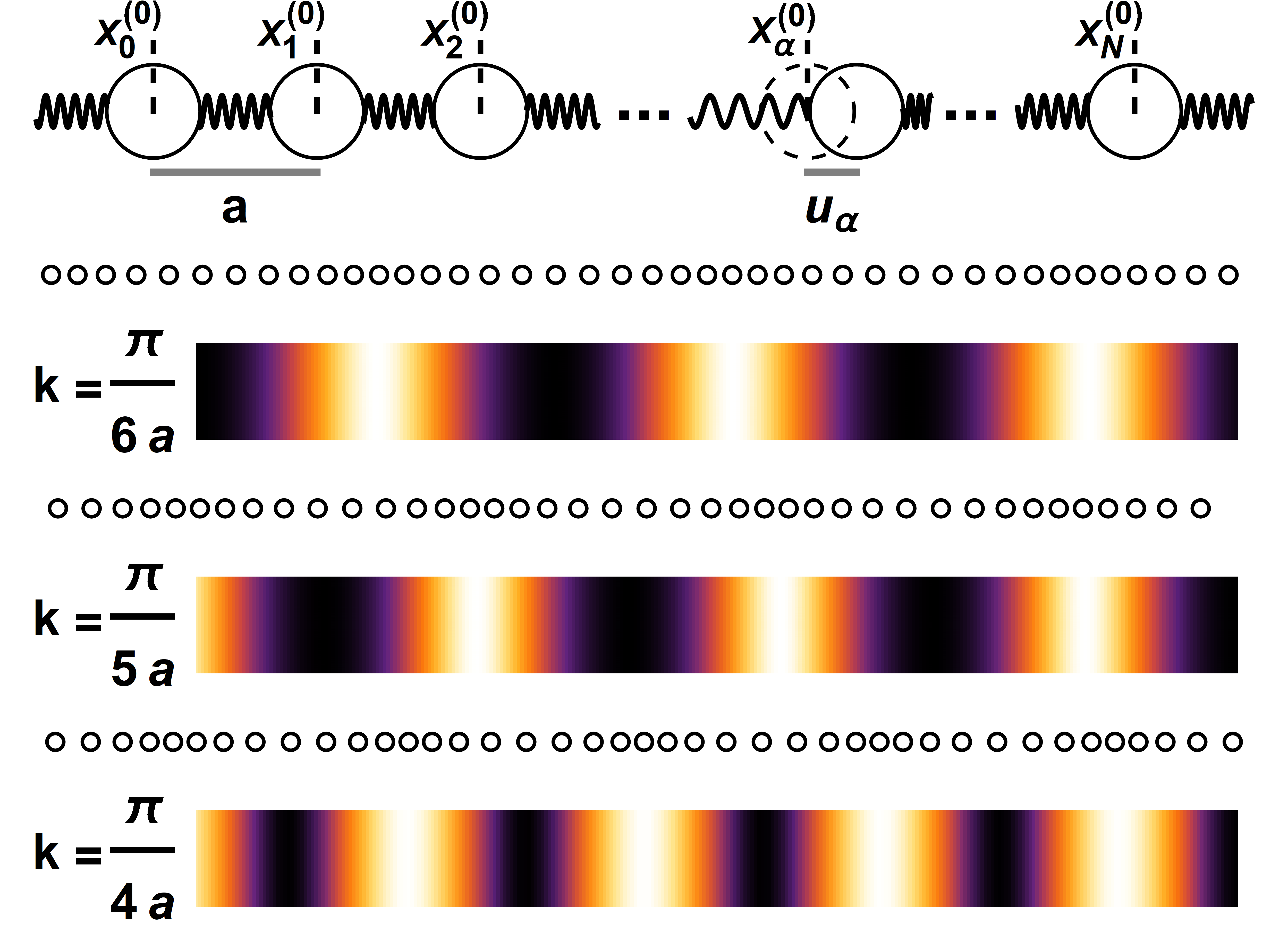}
    \caption{1D Chain of Oscillators. Classical positions of ions and quantum mechanical ionic density distribution for selection of $k$ - phonons. Phonon distortions cause changes in periodicity and electronic band structure.}
    \label{fig:1dChain}
\end{figure}
We claim here that this distortion must be the result of the statistics of the phonons i.e. there must be a sizable population of phonons sitting in a given mode. So, the resulting distortion to the lattice depends entirely on which values of $k$ are deemed more likely by phonon occupation. Additionally, the Kohn anomaly is an apparent distortion to the phonon dispersion relationship which has been shown to be caused by the Fermi surface of electronic states. \cite{polchinski1992effective,hovrava2005stability, kohn1959image, johannes2008fermi}. Likewise this trait of phonon systems is known to have temperature dependence, but the exact dependence is only known qualitatively. We expect these phenomena to be a natural part of the FTMCDFT KS equations, and, given the temperature dependence of the equations, we expect to be able to derive a temperature dependent Peierls gap and Kohn anomaly. This is an attractive application for this framework, because much of the results are already known. We can easily compare these results to alternate methods for confirmation. 
\subsection{KS1N Phonon Dispersion:}
\label{sect:KS1Nphonon}
Since oscillations are supported by molecular bonds, we form a combination of the lattice interaction term, core electron density, and exchange correlation in Eq. $\ref{eq:ksEqnN}$, leaving out the Fermi level conduction electrons, $n_f(x)$. The ionic equation is then rewritten for our purposes like so,
\begin{equation*}
    H_n = \frac{1}{2M}\sum_\alpha\frac{\partial^2}{\partial X^2_\alpha}+V_{latt}(\underline{X})-Z\sum_\alpha\int dx\frac{n_{f}(x)}{\abs{x-X_\alpha}},
\end{equation*}
where $V_{lattice}$ can be as ``real" as one wants. A 3D, real calculation can be performed by replacing $V_{lattice}$ with whatever state of the art potential is desired; however, to make our point, we do not require a model of excessive complexity. For the unperturbed problem, the conduction density will be set to zero and added in during a later step as a perturbation.  We can simply model the 1D lattice with a chain of coupled oscillators, see Fig. \ref{fig:1dChain},
\begin{equation}
    V_{latt}(\underline{X})=\frac{\kappa}{2}\sum^N_\alpha(X_\alpha-X_{\alpha-1})^2,
\end{equation}

\begin{figure}[t!]
    \centering
    \includegraphics[width = 
    \linewidth]{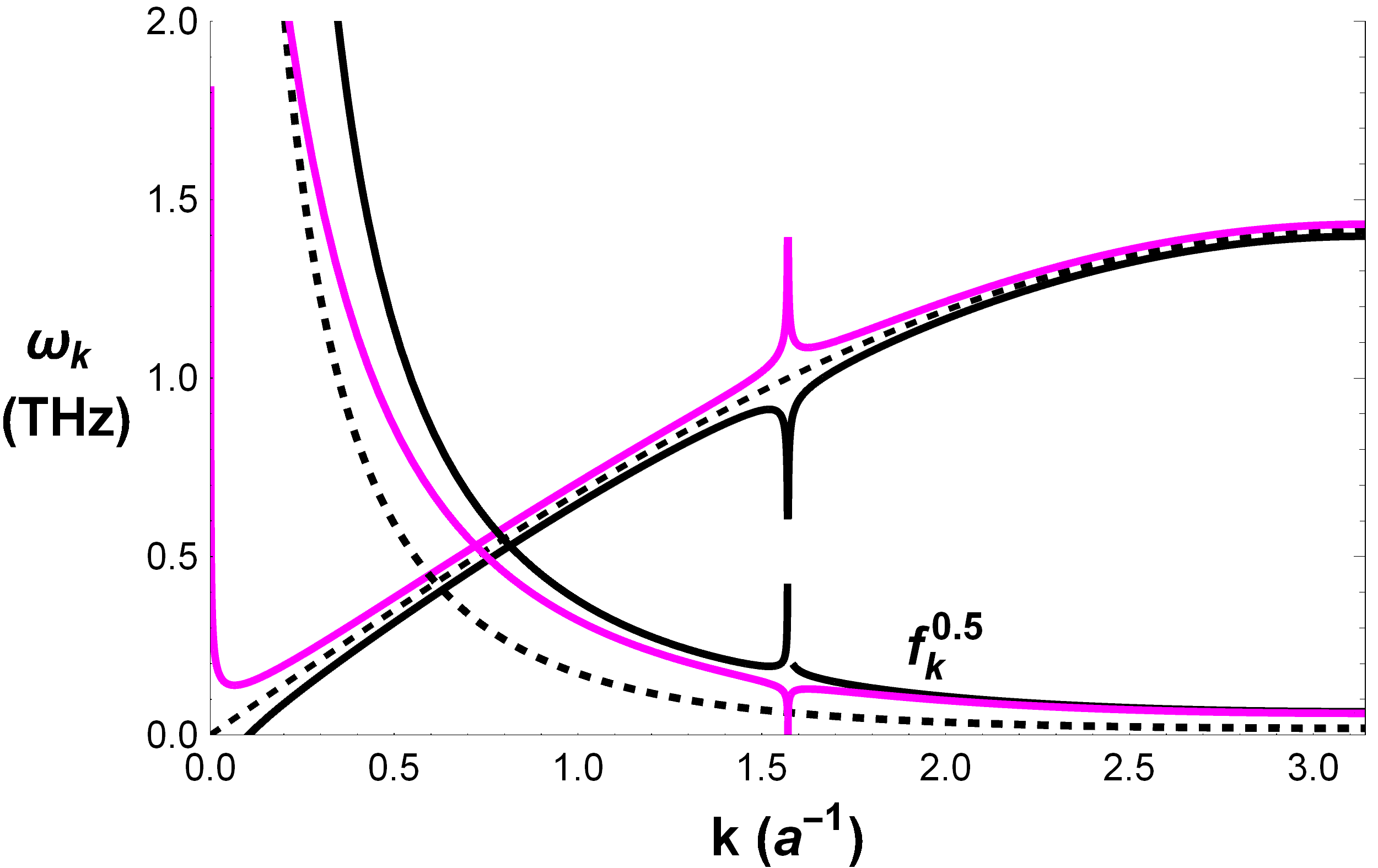}
    \caption{Phonon Dispersion and Occupation. Unperturbed phonon dispersion and occupation are depicted with dashed lines while the low and high energy perturbations are depicted with solid magenta and black lines respectively. These plots are made by setting all constants to unity and using $\frac{\pi}{2a}$ as the representative distorting phonon. However, $k_p$ can take on any value in the BZ which moves the feature to that value. The perturbed dispersion has two singularities for $k=0$ and $k=k_p$ implying a breakdown of the perturbing scheme for those values.}
    \label{fig:phonDisp}
\end{figure}
The procedure for solving for the phonon dispersion curves of this system is well covered in Refs. \cite{ashcroft2022solid, gross1986many}, so only the important points will be recapitulated. The details will be important in the following sections. We assume small deviations, $X_\alpha=X^{(0)}_\alpha+u_\alpha$, and make use of the harmonic approximation, meaning the bare potential is expanded around equilibrium positions and higher order terms are eliminated. Additionally, the constant term is ignored out of lack of interest, and the first order term is argued to be zero because of equilibrium, yielding the general expression,
\begin{equation}
    V\approx\frac{1}{2}\sum^N_{\alpha\beta}u_\alpha u_\beta A_{\alpha\beta}\rvert_{\underline{X}^{(0)}},\hspace{10pt}A_{\alpha\beta}=\frac{\partial^2V}{\partial X_\alpha\partial X_\beta},
\end{equation}
where $A_{\alpha\beta}$ is known as the \textit{force constant matrix}. For full 3D problems, this is a $3\cross 3$ matrix representing the energy change for a given ion when moved in $x,y,z$ neighborhood of its equilibrium position, but, in 1D, it is a scalar. 

The Hamiltonian couples the oscillators adjacently and is obviously nondiagonal. In order to diagonalize, a \textit{rotation} is performed on the classical phase space and degrees of freedom of the system,
\begin{equation}
    Q_k=\frac{1}{\sqrt{N}}\sum_\alpha u_\alpha e^{ikX^{(0)}_\alpha},
    \label{eq:CollCoord}
\end{equation}
which results in a discrete Fourier transform of the entire Hamiltonian. Due to symmetry among the ions, the unperturbed Fourier transform of $A_{\alpha\beta}$ simplifies to,
\begin{equation}
\Lambda_k=\sum^N_\alpha e^{ikX^{(0)}_\alpha}A_{0,\alpha},
\end{equation}
yielding,
\begin{align*}
    \Lambda_k&=4\kappa\sin^2\left(\frac{1}{2}ka\right),\\ \omega_k&=2\sqrt{\frac{\kappa}{M}}\abs{\sin(\frac{1}{2}ka)},
\end{align*}
after some derivatives and trigonometric manipulation. This is plotted as dotted lines with all system parameters set to unity in Fig. \ref{fig:phonDisp}.

\subsection{Nuclear Density \& Phonon Occupation Number}
Using the logic from Sect. \ref{sec:SClogic}, a natural place to search for sudden electro-phonon interaction would be to study the behavior of how phonons populate a given mode and $k$-vector using the Boson occupation number,
\begin{equation}
    f^\beta_k=\{e^{\hbar\beta\omega_k}-1\}^{-1},
\end{equation}
which is immediately known following the dispersion calculation and is plotted in Fig. \ref{fig:phonDisp}. This equation can be stated verbally in the following way. At a given temperature, there will be more phonons with $k$-vector corresponding to lower energy portions of the associated dispersion curve. Since the nuclear density is a statistically averaged collective state \cite{luders2005ab}, the physical manifestation of ionic density that the electrons interact with at a given $\tau$ depends more heavily on these lower energy features. Thus, changes to the energy profile, $\omega_k$, drive the statistical behavior of phonons, and the ionic density profile subsequently changes the electronic density. This is the fundamental idea for our method. 
 
Using phonon operators $b_k$ and ignoring higher energy states, one can construct the ground state harmonic oscillator collective state for each $k$ in the BZ,
\begin{equation*}
    \ket{h_k^0}=\left(\frac{m\omega_k}{\pi\hbar}\right)^{\frac{1}{4}}e^{-\frac{1}{2}\frac{m\omega_k}{\hbar}Q^2_k}.
\end{equation*}
and, with Eq. \ref{eq:gammaGeneral}, we can write the ground state ionic density as \cite{luders2005ab},
\begin{equation}
\label{eq:genGamma}
    \Gamma(\underline{Q})=\sum_{k}f^\tau_{k}\sqrt{\frac{m\omega_k}{\pi\hbar}}e^{-\frac{m\omega_k}{\hbar}Q^2_k}.
\end{equation}
\subsection{Band Structure}
\label{sec:eBand}
We now seek to derive the thermally dependent phonon density's effect on the band structure problem. This is somewhat complicated because the density itself is a sum over many $k$-vector modes in the BZ, each representing a different periodic structure. In normal band structure calculations, one is dealing with a single periodic crystal structure which manifests as a reciprocal lattice of discrete Fourier supports and the electronic BZ. However, in this problem, each phononic quantum state and density yields its own periodic potential that the electrons interact with, so, to address our problem, we must proceed with caution. With that, the KS electronic Hamiltonian looks as follows,
\begin{equation}
H_e=\frac{1}{2m_e}\frac{\partial^2}{\partial x^2}-Z\sum_\alpha\int d \underline{X}\frac{\Gamma(\underline{X})}{\abs{x-X_\alpha}}+\int dx'\frac{n_f(x')}{\abs{x-x'}},
\label{eq:KSelec}
\end{equation}
where we use up the core electron term in the formation of the lattice, so, to introduce a phonon nuclear density, we have already given up the need for an explicit core $n(x)$ term.

Again, setting $n_f(x)=0$ for now, the band structure problem begins by rewriting $H_e$ in the crystal momentum representation,
\begin{equation}
    \bra{k}H_e\ket{k'}\propto\int dxe^{iqx}H_e(x),
    \label{eq:ogFT}
\end{equation}
which is simply a continuous Fourier transform of the electronic Hamiltonian with respect to $q = k-k'$. At this stage, $q$ is a continuum variable, but one should expect a constraint in the following math which places this parameter on a reciprocal lattice. Usually, one only deals with the reciprocal lattice related to the equilibrium positions of the ions. This reciprocal lattice determines the location of gaps which occur in the band structure of electrons. However, in this work and given the nature of Peierls phenomenon, we deal with the sudden opening of gaps which are located at places not coinciding with the typical reciprocal lattice. These gap openings are instead related to a new periodic structure imposed by phonon distortions. For this reason, it is absolutely necessary to be pedantic in these details. 

The transformation of the kinetic energy term in Eq. \ref{eq:KSelec} is straightforward if one accepts foundational quantum mechanics as straightforward, $\nabla^2\rightarrow q^2$. However, the ionic term needs special attention. Since the Fourier transform Eq. \ref{eq:ogFT} only concerns the electronic coordinate, $x$, the transform can be passed straight through to the Coulombic term,
\begin{equation*}
    V_q=Z\sum_\alpha\int d\underline{X}\Gamma(\underline{X})\bra{k}\frac{1}{\abs{x-X_\alpha}}\ket{k'}.
\end{equation*}
We can pull an exponential of the ionic coordinate out using the Fourier shift theorem,
\begin{equation*}
    V_q=Z\sum_\alpha\int d\underline{X}\Gamma(\underline{X})e^{-iqX_\alpha}\bra{k}\frac{1}{\abs{x}}\ket{k'},
\end{equation*}
and we use a well known solution to the Fourier transform of the Coulomb interaction, $\frac{1}{\abs{x}}$,
to write,
\begin{equation*}
    V_q=\frac{4\pi Z}{q^2}\sum_\alpha\int d\underline{X}\Gamma(\underline{X})e^{-iqX_\alpha}.
\end{equation*}
Finally, one expands the ionic term in the complex exponential using the small deviation assumption from the previous section,
\begin{equation}
\label{eq:ogAugFunc}
    V_q=\frac{4\pi Z}{q^2}\sum_\alpha e^{-iqX^{(0)}_\alpha}\int d\underline{u}\Gamma(\underline{u})e^{-iqu_\alpha},
\end{equation}
leaving behind a partial Fourier transform of the ionic density with respect to one of the ionic coordinates indexed by $\alpha$. 

We pause here to briefly comment on the structure of Eq. \ref{eq:ogAugFunc}. Given some arbitrary periodic lattice potential $f(x)$, one can demonstrate that the Fourier transform will contain a prefactor which is a discrete sum of plane waves, 
\begin{equation*}
    \tilde{f}(q)=\sum^N_\alpha e^{-iqX^{(0)}_\alpha}\int dx f(x)e^{-iqx}.
\end{equation*}
In the limit as $N\rightarrow\infty$,
\begin{equation}
    \sum^N_\alpha e^{-iqX^{(0)}_\alpha}\propto\sum_{K\in\mathcal{L}^*} \delta(q-K),
    \label{eq:recLatt}
\end{equation}
where $\mathcal{L}^*$ is the typical reciprocal lattice. It is thus the appearance of this prefactor which provides the expected restriction on $q$. To see this, one can perform an inverse Fourier transform on $\tilde{f}(q)$. The sum of deltas within the integral will have the effect of picking out values of $q$ such that $q=K$. Given the periodic nature of our problem, this sum of plane waves appears for us in Eq. \ref{eq:ogAugFunc} as well; however, it is augmented an integral of the ionic density multiplied by a plane wave of a single ionic coordinate. This indicates that the presence of phonons changes the character of the reciprocal lattice which is exactly what we expect to be the case for lattice distortions.

Using Eq. \ref{eq:genGamma}, we can write out the integral explicitly as,
\begin{equation}
\small
    V_q=\frac{4\pi Z}{q^2}\sum_{k}f^{\tau}_{k}\sqrt{\frac{m\omega_{k}}{\pi \hbar}}\sum_\alpha e^{-iqX^{(0)}_\alpha}\int d\underline{u}e^{-\frac{m\omega_k}{\hbar}Q^2_{k}}e^{-iqu_\alpha}.
    \label{eq:bandTrans1}
\end{equation}
which we now wish to evaluate by replacing the $Q_k$ coordinate with the real part of its definition in Eq. \ref{eq:CollCoord}. The reason that a real part is used is due to the fact that the integrand originates from a necessarily real valued density function. Furthermore, viewing the classical phase space as a vector space, we are free to write the rotation made in Eq. \ref{eq:CollCoord} in matrix and vector notation as $\textbf{Q}_k=\text{Re}\{\textbf{M}_k \textbf{u}$\}, where $\textbf{M}_k$ is a rotational matrix of complex exponentials, the $\textbf{Q}_k$'s locate a point along the new phonon coordinate basis vectors, and $\textbf{u}$ is a vector of real valued ionic coordinates in phase space. Since $Q^2_k$ is meant to be interpretted as magnitude from the origin along this coordinate axis in the prequantization bundle, we write $Q_k^2=\abs{\textbf{Q}_k}^2$. Here, $\abs{\hspace{2pt}\cdot\hspace{2pt}}$ is the standard vector norm. In other words, to take the square of this coordinate, we take the sum of the components squared like so,
\begin{equation*}
    Q_k^2=\frac{1}{N}\sum_\alpha u_\alpha^2 cos(kX^{(0)}_\alpha)^2.
\end{equation*}
\begin{widetext}
With that, we expand the $Q^2_k$ within the Gaussian exponential yielding, 
\begin{equation}
V_q=\frac{4\pi Z}{q^2}\sum_{k}f^{\tau}_{k}\sqrt{\frac{m\omega_{k}}{\pi \hbar}}\sum_\alpha e^{-iqX^{(0)}_\alpha}\int d\underline{u}e^{-iqu_\alpha}e^{-\frac{m\omega_k}{N\hbar}\sum_{\alpha'} u^2_{\alpha'} cos^2(kX^{(0)}_{\alpha'})}\nonumber.\\
\end{equation}
The exponential can be equivalently written as a product in the following way,
\begin{equation}
    V_q=\frac{4\pi Z}{q^2}\sum_{k}f^{\tau}_{k}\sqrt{\frac{m\omega_{k}}{\pi \hbar}}\sum_\alpha e^{-iqX^{(0)}_\alpha}\int d\underline{u}e^{-iqu_\alpha}\prod_{\alpha'}e^{-\frac{m\omega_k}{N\hbar}u^2_{\alpha'} cos^2(kX^{(0)}_{\alpha'})}\nonumber.
\end{equation}
The integral is taken over all ionic coordinates, so we can separate every $u_{\alpha'}$ exponential within the product into its own integral like so,
\begin{align}
V_q=\frac{4\pi Z}{q^2}\sum_{k}f^{\tau}_{k}\sqrt{\frac{m\omega_{k}}{\pi \hbar}}\sum_\alpha e^{-iqX^{(0)}_\alpha}\underbrace{\int du_1 e^{-\frac{m\omega_k}{N\hbar}u^2_{1} cos^2(kX^{(0)}_{1})}}_{=1}...&\underbrace{\int du_\alpha e^{-\frac{m\omega_k}{N\hbar}u^2_{\alpha} cos^2(kX^{(0)}_{\alpha})}  e^{-iqX^{(0)}_\alpha}}_{\propto e^{-\frac{\pi^2 N\hbar}{2m\omega_k}cos(kX^{(0)}_\alpha)^{-2}q^2}}\nonumber\\
&...\underbrace{\int du_N e^{-\frac{m\omega_k}{N\hbar}u^2_{N} cos^2(kX^{(0)}_{N})}}_{=1},
\label{eq:tinyFT}
\end{align}
where the unprimed $u_\alpha$ from the original Fourier transform is brought into the $\alpha$ integral as well. Thus, one is left with single Gaussian Fourier transform with respect to the unprimed $\alpha$ coordinate and $(N-1)$ Gaussian integrals with respect the the leftover coordinates.
\end{widetext}

Since these are Gaussian densities, every integral in Eq. \ref{eq:tinyFT} must be normalized, so all that remains is the result of the partial $u_\alpha$ Fourier transform, yielding the total expression for $V_q$,
\begin{equation}
    V_{q}=\frac{4\pi Z}{q^2}\sum_{k}f^{\tau}_{k}\sum_\alpha e^{-iqX^{(0)}_\alpha}g_{k,\alpha}(q).
\label{eq:matrixElem}
\end{equation}
where,
\begin{equation*}
g_{k,\alpha}(q) \propto e^{-\frac{\pi^2 N\hbar}{2m\omega_k}sec(kX^{(0)}_\alpha)^{2}q^2},
\end{equation*}
is the result of the $u_\alpha$ Fourier transform. This quantity drives the effect on the Fourier supports for the potential. Whenever $kX^{(0)}=\pi/2$, the secant function is singular, forcing the exponential to zero. Therefore, $g_{k,\alpha}(q)$ has the effect of deleting elements of the plane wave sum which the electrons experience as a change to Eq. \ref{eq:recLatt}, the reciprocal lattice, and the BZ. We denote the augmented discrete plane wave sum as $\Delta_k(q)$ and the new reciprocal lattice as $(\mathcal{L}')^*$. With this understanding, we condense the matrix element for the band structure problem like so
\noindent
\begin{equation}
\label{eq:finalMatrixElem}
    V_q =\frac{4\pi Z}{q^2}\sum_{k}f^{\tau}_{k}\Delta_{k}(q),
\end{equation}
to emphasize the role of the statistical averaging procedure with respect to the determination of the distorted periodicity of the lattice, concluding the derivation.

\begin{figure}[t!]
    \centering
    \includegraphics[width=\columnwidth]{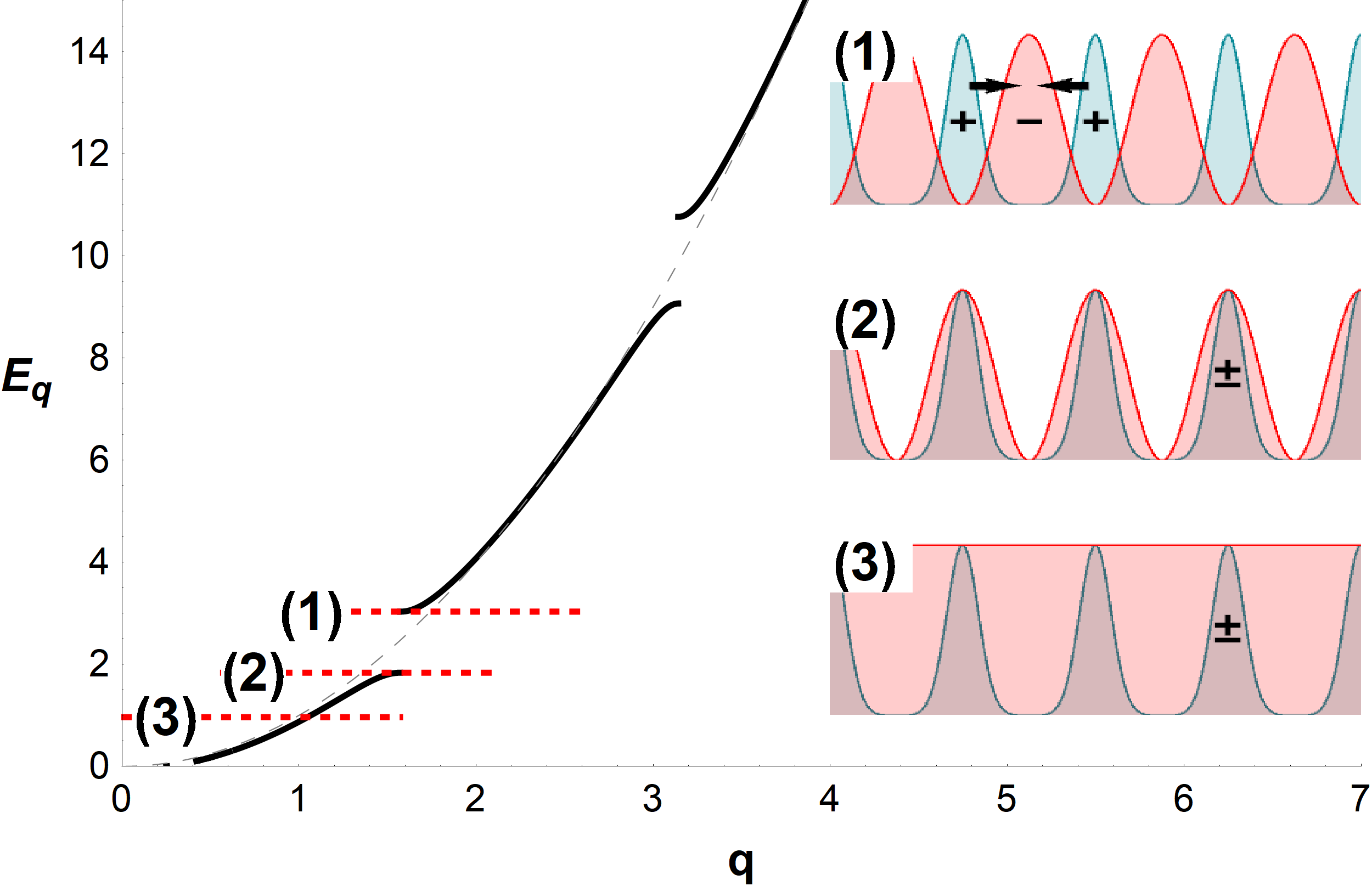}
    \caption{Conduction Density. The conduction electronic density reaction to the Peierls gap takes on three possibilities depending on location of Fermi level. The inset depicts a general ionic, positive distribution (blue) and the conduction density of each possibility (red). The higher energy conduction density accumulates in between the phonon distorted ionic density which has the capability of re enforcing the collective oscillation.}
    \label{fig:PeierlsGapEband}
\end{figure}

These new supports have the known effect of introducing matrix elements to the problem which were previously zero for a non-distorted system, and the introduction of these matrix elements opens a gap for electrons as depicted in Fig. \ref{fig:PeierlsGapEband}. This is of course a restatement of Peierls phenomena. However, our perturbative prescription successfully retains the statistical generality of this simple model in the form of the Boson occupation number in Eq. \ref{eq:matrixElem}, and the gap opening is allowed for any $k$-phonon distortion deemed most likely by statistics.

Additionally, in Fig. \ref{fig:PeierlsGapEband}, we note that there are three possibilities of electronic conduction density response to a $k_p$ - phonon distortion: 1) Supra gap: $n_f\propto \cos(\frac{1}{2}k_p x)^2$, 2) Sub gap: $n_f\propto \sin(\frac{1}{2}k_p x)^2$, and 3) Mid band: $n_f\propto 1$. The negative charge distribution that emerges from the distortion will depend entirely on where the Fermi level is with respect to the phonon gap opening. In typical Peierls phenomenon, the phonon distortion is fixed at $k_p=\pi$ which is known as the dimerization of the 1D lattice. This distortion provokes a band gap opening at exactly $\frac{\pi}{2}$. If every ion provides one valence electron, the Fermi surface will be a discrete set of two points at $q = \pm\frac{\pi}{2}$. So, scenario 1 is not possible, and we are left with scenario 2 where the Fermi level is in between the gap, making the material an insulator. However, because our work leaves the distorting $k_p$ to be determined by the distribution $f^\tau_k$, the gap opening can be shifted in relation to the Fermi surface in a way which could satisfy scenarios one and three as well.

\subsection{Kohn Anomaly Derivation}
The three possible electronic densities responses are now introduced to the problem as perturbations where we seek to derive the electronic densities effect on the phonon statistical environment. Returning to the ionic problem from Sect. \ref{sect:KS1Nphonon} once again, we simplify the conduction density term by approximating the Coulomb term with a Dirac delta like so,
\begin{align*}
    \sum_\alpha\int dx\frac{n_{f}(x)}{\abs{x-X_\alpha}}&\approx\sum_\alpha\int dxn_{f}(x)\delta(x-X_\alpha)\\
    &=\sum_\alpha n_{f}(X_\alpha).
\end{align*}
We now must follow a similar procedure to the original phonon dispersion calculation; however, there are some steps that require attention because of the alternate periodic nature of the perturbation. Starting from the Harmonic approximation, we split the force constant matrix into the original and perturbation contribution, $V=V_{latt}+V_{f}$. This split yields the original dispersion relationship for the unperturbed problem with the addition of,
\begin{equation*}
    V_f=\frac{1}{2}\sum^N_{\alpha\beta}u_\alpha u_\beta \frac{\partial^2}{\partial X_\alpha\partial X_\beta}\sum_{\alpha'} n_{f}(X_{\alpha'})\rvert_{\underline{X}^{(0)}}.
\end{equation*}
This term has three summations, but, after differentiating twice, only one remains,
\begin{equation}
    V_f=\frac{1}{2}\sum^N_{\alpha}u^2_\alpha \frac{\partial^2}{\partial X^2_\alpha} n_{f}(X_{\alpha})\rvert_{\underline{X}^{(0)}}
    \label{eq:step1}
\end{equation}
Evaluating the twice derivative with respect to the three density responses, mid gap, sub gap, and supra gap, yields zero and $\pm\frac{Zk^2_p}{2}\cos(k_pX^{(0)}_\alpha)$ respectively. This perturbation obviously does not share the periodic structure of the equilibrium lattice because the cosine function is a periodic function of $X^{(0)}_\alpha$ parametrized by the phonon distorting $k_p$, which inherently possesses alternate periodicity. Placing this quantity into Eq. \ref{eq:step1}, we expand $u_\alpha$ in terms of the coordinates $Q_k$ using the reverse of Eq. \ref{eq:CollCoord},
\begin{equation*}
    V_f=\pm\frac{Zk^2_p}{4}\sum_{kk'}Q_{k}Q_{k'}\frac{1}{N}\sum^N_{\alpha}e^{-i(k+k')X^{(0)}_\alpha}\cos(k_pX^{(0)}_\alpha).
\end{equation*}
We then expand the cosine function in terms of complex exponentials, and, when combined with the $\alpha$-sum of exponentials, one can use completion to write the sum as Kronecker deltas,
\begin{equation*}
V_f=\pm\frac{Zk^2_p}{2}\sum_{kk'}Q_{k}Q_{k'}\{\delta_{k',k_p-k}+\delta_{k',-k_p-k}\}.
\end{equation*}
These can be evaluated to remove one of the summations yielding a perturbation to the collective coordinate simple harmonic oscillators,
\begin{equation}
\label{eq:vfQrep}
    V_f=\pm\frac{Zk^2_p}{2}\sum_{k}\{Q_{k}Q_{k_p-k}+Q_{k}Q_{-k_p-k}\}.
\end{equation}
Now, letting $\varepsilon_k^{(0)}=\frac{1}{2}\hbar\omega_k$ be the unperturbed energy proportional to the dispersion relationship calculated in \ref{sect:KS1Nphonon}, we seek out the first order perturbation resulting from Eq. \ref{eq:vfQrep}, $\varepsilon_k^{(1)}=\bra{\Phi_0}V_f\ket{\Phi_0}$, which is simply the expectation value of the conduction electron density perturbation with respect to the ground state of the unperturbed system.
\begin{widetext}
Since the system is uncoupled with respect to the $Q_k$ coordinate, $\ket{\Phi_0}=\bigotimes_{k}\ket{h_{0,k}}$ is a seperable ground state of the independent phononic oscillators, and the expectation value of Eq. \ref{eq:vfQrep} is,
\begin{equation}
    \varepsilon_k^{(1)}=\pm\frac{Zk^2_p}{2}\bigotimes_{k'}\bra{h_{0,k'}}\{Q_{k}Q_{k_p-k}+Q_{k}Q_{-k_p-k}\}\bigotimes_{k''}\ket{h_{0,k''}}.\label{eq:line1}
\end{equation}

The tensor product in Eq. \ref{eq:line1} contains many states with respect to $Q_k$ that are not present within the perturbation. This can be seen directly from Eq. \ref{eq:vfQrep}, since only one $k$-phonon parameter appears. For this reason, any $\ket{h_{0,k'(k'')}}$ such that $k',k''\neq k$, pass directly through the perturbation term resulting in a unitary inner product. Therefore, one is left with two terms for each $Q_{\pm k_p-k}$,
\begin{equation}
    \varepsilon_k^{(1)}=\pm\frac{Zk^2_p}{2}\{\bra{h_{0,k}}Q_{k}\ket{h_{0,k}}\bra{h_{0,k_p-k}}Q_{k_p-k}\ket{h_{0,k_p-k}}+\bra{h_{0,k}}Q_{k}\ket{h_{0,k}}\bra{h_{0,-k_p-k}}Q_{-k_p-k}\ket{h_{0,-k_p-k}}\}\label{eq:line2}
\end{equation}
Since each $\ket{h_{0,k}}$ is a Gaussian function of $Q_k$, these leftover inner products within Eq. \ref{eq:line2} are integrals of the form $\int dx \hspace{2pt}x e^{-\sigma x^2}$
which can be solved using an integral table,
\begin{equation}
\varepsilon_k^{(1)}=\pm\frac{Zk^2_p}{2\pi\sqrt{\omega_k}}\{\frac{1}{\sqrt{\omega_{k_p-k}}}+\frac{1}{\sqrt{\omega_{-k_p-k}}}\},\label{eq:line3}
\end{equation}
concluding the derivation for the first order perturbation. 
\newpage
\end{widetext}

Eq. \ref{eq:line3} is plotted in Fig. \ref{fig:phonDisp} alongside the unperturbed dispersion and occupation, where the high energy perturbation (solid black) reveals behavior that is markedly similar to the Kohn anomaly.
The low energy perturbation (magenta) reveals an inverted version of the Kohn anomaly-like plot; however, this perturbation is non-physical, because, while our derivation of the Peierls phenomena in Sect. \ref{sec:eBand} retains all statistical generality, this perturbative derivation does not. The line of thinking presented above makes the prior assumption that the phonon occupation has already developed features which dramatically shift the average to a value of $k=k_p$ far from the origin so that there is absolute statistical certainty that the system is in that particular mode of oscillation. This fact is on full display in Fig. \ref{fig:phonDisp} where the higher energy perturbed phonon occupation must have a delta like spike associated with the singularity. For this scenario, the system requires less energy to produce a $k_p$ - phonon which subsequently makes it a more likely style of oscillation. The lower energy perturbed phonon occupation implies the opposite that there are no phonons in the $k_p$ mode. However, if there are no $k_p$-phonons present then the perturbation didn't exist in the first place.

Essentially, we assume that the system possesses qualities that encourage the production of $k_p$ phonons, and the electrons fall into a configuration which either does or does not facilitate those vibrations. Specifically, the high energy perturbation causes conduction electrons to collect in between the ionic high density regions which re enforces the oscillation, allowing us to make the assumption mentioned above. The low energy density does not have this effect since the electrons locate themselves on top of the high ionic density regions, invalidating the assumption, and we find nonphysical results. Additionally, this perturbative technique only applies for $k\neq 0, k_p$. At these values of $k$, the perturbation is singular i.e. very large, meaning some other form of analysis would be needed to derive exact values for these points. In actuality, there is a fundamental distinction between Eq. \ref{eq:line3} and the true Kohn anomaly due to our methods of derivation. In order to establish a stronger link between the two, one would need to make use of a non local Coulomb interaction in contrast to what we have done in this work. 

Even still, our line of reasoning firmly establishes a relationship between the phonon statistics and the electron density response. As a continuation, we know from experiments that the Kohn anomaly for the 1D chain is temperature dependent and non-singular for temperatures above critical, $\tau_C$ \cite{zhu2017misconceptions}. Because of this derived relationship, one could imagine that a concave up, lower energy region of the phonon dispersion curve would accumulate more phonons at any arbitrary $\tau> \tau_C$, but, because the system has all types of phonons present in large numbers, this fact makes no difference to the electrons which are presented with a randomly vibrating lattice. However, as $\tau\rightarrow\tau_C$, there would be a momentary shift in the phonon occupational average that the valence electrons react to. In that case, this average would be heavily influenced by these low energy, high occupation portions of the dispersion curve. Thus, the functional form of the Kohn anomaly develops a temperature dependent term appearing in the denominator of our first order correction which makes the $f_{k=k_p}$ finite. These functional dependencies could be derivable from this scheme as well. 

\section{Conclusion}
First principles, theoretically predictive techniques for SC properties are a key component for the search and synthesis of SC materials. Macroscopically coherent systems currently possess a certain amount of mystique which stems from an insufficient understanding of the underlying quantum machinery. This work is a small part of a large effort to shift fundamental many-body theory by providing a direct path from the many-body Hamiltonian. 

To that end, referring to Sect. \ref{sec:SClogic}, we have presented here a novel first principles technique which firmly connects the behavior of a materials phonon statistics with the accumulation of electronic density for a 1D system. This techinique is successfully used here to derive fundamental electro-phonon interactive phenomena, retaining statistical generality throughout. Using the logic within this work, we have demonstrated that in this MCDFT KS scheme the phenomena of Peierls effect and Kohn Anomaly do not require special treatment to derive. These two features of an exceedingly simple 1D system are electro-phonon phenomena which are directly predicted, and their appearance motivates us to continue this line of reasoning. It is hopeful then that using this scheme, one could possibly directly predict the SC nature that stems from phonon mediation.

One could continue into the 2D and 3D realm, and, because of the generalized nature of the preliminary sections, one could choose a lattice potential which is real enough by including state of the art potentials and exchange correlation. From there, a powerful computational scheme could be used. Additionally, it is known that these features derived here are temperature dependent, so further work can be done for 1D systems to derive such dependencies.
\bibliographystyle{apsrev4-2}
\bibliography{main}
\end{document}